\documentclass[pss]{wiley2sp} 
\usepackage{amsmath}
\usepackage{amsfonts}
\usepackage{amssymb}
\usepackage{graphicx}
\usepackage{color}

\definecolor{red}{rgb}{1,0,0}
\definecolor{blue}{rgb}{0,0,1}

\begin{document}

\title{Towards combined transport and optical studies of the 0.7-anomaly in a quantum point contact}

\titlerunning{Towards combined transport and optical studies of the 0.7-anomaly in a QPC}

\author{%
  Enrico Schubert\textsuperscript{\Ast,\textsf{\bfseries 1}},
  Jan Heyder\textsuperscript{\textsf{\bfseries 1,2}},
  Florian Bauer\textsuperscript{\textsf{\bfseries 1,2}},
  Wolfgang Stumpf\textsuperscript{\textsf{\bfseries 3}},
  Werner Wegscheider\textsuperscript{\textsf{\bfseries 3}},
  Jan von Delft\textsuperscript{\textsf{\bfseries 1,2}},
  Stefan Ludwig\textsuperscript{\textsf{\bfseries 1}},
  Alexander H\"ogele\textsuperscript{\textsf{\bfseries 1}}
  }

\authorrunning{E. Schubert et al.}

\mail{e-mail
  \textsf{enrico.schubert@physik.uni-muenchen.de}, Phone:
  +49-89-2180-3349, Fax: +49-89-2180-3182}

\institute{%
  \textsuperscript{1}\,Center for NanoScience and Fakult\"at f\"ur Physik,\\
Ludwig-Maximilians-Universit\"at M\"unchen,
Geschwister-Scholl-Platz 1, 80539 M\"unchen, Germany\\
  \textsuperscript{2}\,Arnold Sommerfeld Center for Theoretical Physics, \\
Ludwig-Maximilians-Universit\"at M\"unchen,
Theresienstrasse 37, 80333 M\"unchen, Germany\\
  \textsuperscript{3}\,Laboratory for Solid State Physics, ETH Z\"urich, CH-8093 Z\"urich,
    Switzerland
  }

\received{XXXX, revised XXXX, accepted XXXX} 
\published{XXXX} 


\abstract{%
%
%
%
\abstcol{%
A Quantum Point Contact (QPC) causes a one-dimensional
constriction on the spatial potential landscape of a two-dimensional electron system. By tuning the voltage applied on
a QPC at low temperatures the resulting regular step-like electron
conductance quantization can show an additional kink near pinch-off
around $0.7e^2/h$, called $0.7$-anomaly. In a recent publication, 
we presented a combination of
theoretical calculations and transport measurements that lead to a detailed
understanding of the microscopic origin of the 0.7-anomaly. 
Functional Renormalization Group-based calculations were performed exhibiting the $0.7$-anomaly
even when no symmetry-breaking external magnetic fields are
involved. According to the calculations the electron spin\linebreak ~\vspace{-5mm}}{
susceptibility is enhanced within a QPC that is tuned in the region of the 0.7-anomaly. Moderate
externally applied magnetic fields impose a corresponding
enhancement in the spin magnetization. In principle, it should be possible to map out this spin
distribution optically by means of the Faraday rotation technique. Here we report the
initial steps of an experimental project aimed at realizing such measurements.
Simulations were performed on a particularly pre-designed
semiconductor heterostructure. Based on the simulation results a
sample was built and its basic transport and optical properties 
were investigated. Finally, we introduce a sample gate design, suitable for combined
transport and optical studies. }}

%
%


\maketitle   






\section{Introduction}
A quantum point contact (QPC) is a short, 1-dimensional constriction
usually realized within a 2-dimensional electron system (2DES), by
applying voltage to metallic gates, thereby depleting the electrons
beneath and only leaving a narrow transport channel whose width can be
tuned by the applied gate-voltage. When a QPC is opened up by changing
the applied gate-voltage, its conductance not only, famously, rises in
integer steps of the conductance quantum, $G_Q = 2 e^2/h$
\cite{Wees1988,Wharam1988,Buttiker1990}, but also shows a
shoulder-like intermediate step at the onset of the first plateau,
around $\simeq 0.7 G_Q$, that has a very intriguing dependence on
temperature ($T$), magnetic field ($B$) and source-drain voltage
($V_\mathrm{SD}$)
\cite{Thomas1996,Thomas1998,Appleyard2000,Cronenwett2002}.  This
phenomenon is known as the 0.7-anomaly.  A succinct summary of the
status of various previous theoretical treatments thereof may be found
in \cite{Micolich2011}.

In a recent publication \cite{Bauer_Nature_2013}, we presented a
combination of theoretical calculations and transport measurements
that lead to a detailed understanding of the microscopic origin of the
0.7-anomaly. It is caused by a smeared van Hove peak in the
local density of states (LDOS), whose weight, shape and position
depends on sample geometry (width, length and shape of the QPC
confinement potential). The peak enhances the effect of interaction by
two main mechanisms: first, it enhances the effective Hartree barrier,
and thus the elastic back-scattering due to Coulomb repulsion; second,
inelastic scattering is enhanced once phase space is opened up by
increasing the temperature or the source-drain bias voltage.

The present paper serves two purposes. First, in section 2 we
summarize some of the main results from \cite{Bauer_Nature_2013},
highlighting, in particular, one of its central predictions: the local
spin susceptibility is predicted to be anomalously enhanced in the
vicinity of the QPC.  Second, in section 3 we describe the initial
stages of an experimental project that ultimately aims at 
detecting the predicted anomalous behavior of the spin susceptibility
in a QPC by optical methods.

\begin{figure}[b]
\centering
\includegraphics[width=7 cm]{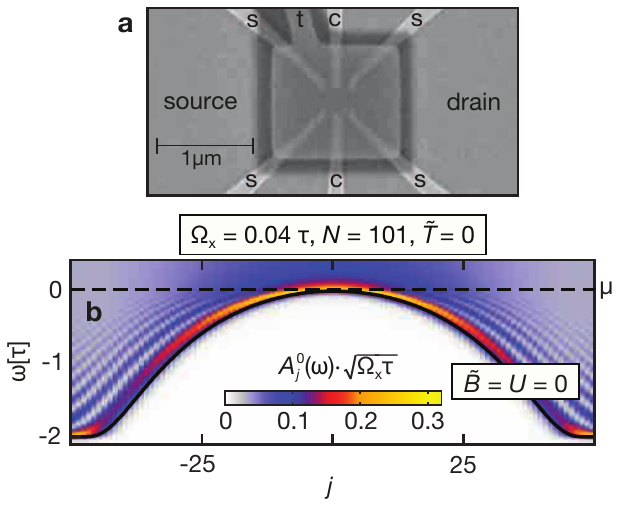}
\vspace{1mm}
\caption{\textbf{a}, gate layout of one of our samples designed to investigate the
0.7-anomaly of QPCs. The metal gates (light gray) are placed with the help of
electron-beam lithography on the surface of GaAs\,/\,AlGaAs heterostructures. The
darker shaded areas consists of cross-linked PMMA which is used to electrically
isolate vertically stacked metal gates. A semi-transparent titanium top-gate (gray,
on top of PMMA) covers the nanostructure including its leads. Experiments were done
at ultra-low temperatures (base temperature of 17\,mK, electron temperature $T_0
\simeq $ 30\,mK). \textbf{b}, the bare local density of states, $A_j^0(\omega)$ , in
the central region of the QPC as a function of site $j$ and frequency $\omega$. The
maximum of $A_j^0$ follows the shape of the band, i.e. the shape of the applied
potential, resulting in a distinct ridge-like structure (yellow), the van Hove ridge.} \label{Fig:sample}
\end{figure}

\section{Microscopic origin of the 0.7-anomaly}

We use a multi-gate sample that gives us direct control over the geometry of the
confinement potential, defining the QPC. In this section we present both theoretical calculations and
experimental measurements of transport of the lowest transverse mode, finding very
good qualitative agreement for the conductance as a function of applied gate voltage
at both zero and finite magnetic field. We predict the shapes of the spin-resolved
conductance curves and show that the 0.7-anomaly coming out of our calculations is directly linked to a
smeared van Hove singularity, a maximum in the LDOS, located at the top of the 1D-potential which, in
combination with interactions, gives rise to a strongly enhanced spin-susceptibility.
We show that the strength of interaction within the constriction can be tuned by a
global top-gate.

Our QPC design (see FIG.\ref{Fig:sample}\textbf{a}) allows a detailed tuning of the
confinement potential within the 2DES both along and
perpendicular to the
electron propagation direction through the QPC, thereby defining the length and width of the QPC respectively . In addition to the two central gates (c) and
four side gates (s), allowing for a fine-tuning of the effective 1D barrier, the
sample also contains a global top-gate (t) to adjust the charge carrier density of
the 2DES. The advantages of such a sample geometry are: first, in experiments
it is often difficult to clearly distinguish the 0.7-anomaly from unwanted disorder
related resonances, which can alter the results in an uncontrolled way. In our case
the enhanced tunability via multiple control gates
facilitates an unambiguous identification of the 0.7-anomaly and a sufficient
separation from disorder induced resonances. This is extremely important for a
quantitative comparison with model calculations, which assume a potential landscape
without disorder. Second, the multi-gate structure enables us to monitor the evolution
of the QPC properties with varying length and width independently, which makes it a
versatile tool for a systematic analysis of the 0.7-structure.

The experimental realization of a QPC is modeled by a simple potential barrier describing the effective 1D-potential along the electronic transport direction. Information about the transverse structure of the channel is fully incorporated into a space-dependent model parameter $U$, defining the strength of interactions. After discretizing space the model Hamiltonian is given by 

\begin{eqnarray}
\label{eq:model}
\hat{H} = \sum_{j\sigma} \left[
E_{j\sigma} \hat n_{j\sigma}
- \tau_j ( d^\dagger_{j+1 \, \sigma} d_{j \sigma} + \rm{h.c.}) \right] + \notag \\
+ \sum_j U_j \hat{n}_{j\uparrow} \hat{n}_{j \downarrow} \; . \nonumber \\
\hspace{-5mm} \phantom{.}
\end{eqnarray}

It describes an infinite tight-binding chain with nearest-neighbor hopping $\tau_j$,
on-site interactions $U_j$, and a uniform magnetic field, $\tilde{B} = g
\mu_\textrm{B}B $, acting only to Zeeman-split opposite spins. (we
use symbols with or without tildes, e. g. $\tilde{B}$ or $B$, to distinguish model
parameters from experimental ones, respectively). Orbital effects are
neglected, a good approximation if the field is parallel to the two-dimensional
electron system). The on-site energy, $ E_{j,\sigma} = E_j -\frac{\sigma}{2}
\tilde{B}$, in combination with the hopping, $\tau_j$, both vary smoothly with $j$,
thus creating an effective potential barrier $V_j = E_j - (\tau_j + \tau_{j+1}) -
\mu$, measured w.r.t. the chemical potential, $\mu$ (we use $\mu=0$). We choose
$U_j\neq0$ and $E_j\neq 0$ only for $N$ sites, symmetric around $j=0$ that define the
extent of the QPC. $U_j$ is constant in the center of the QPC with $U_0=U$ and drops
smoothly to zero for $|j|\to N/2$. We choose the potential, $V_j$, to be symmetric
and parabolic near the top, $V_j = \tilde{V}_{\rm c} - \Omega_x^2/(4\tau_0) j^2$  with barrier-height $\tilde{V}_{\rm
c}$, mimicking the role of gate voltage from experiment, and curvature $\Omega_x$,
defining the effective length of the QPC (see supplementary information of
\cite{Bauer_Nature_2013} for more details). FIG.\ref{Fig:sample}\textbf{b} shows the
bare LDOS, $A_j^0(\omega)$, of the QPC as a function of site $j$ and frequency
$\omega$. The LDOS has a maximum right above the band bottom, visible as a yellow-red
structure, that follows the shape of the potential (black thick line). This
structure, which lies at the heart of the explanation for the 0.7-anomaly, will be
called a ``van Hove ridge''. The ridge maximum lies slightly higher in energy than
the potential $V_j$, by an amount that scales like the potential curvature
$\Omega_x$, where the LDOS is proportional to $1/\sqrt{\Omega_x \tau_0}$. 

\begin{figure}[t]
\centering
\includegraphics[width=\linewidth]{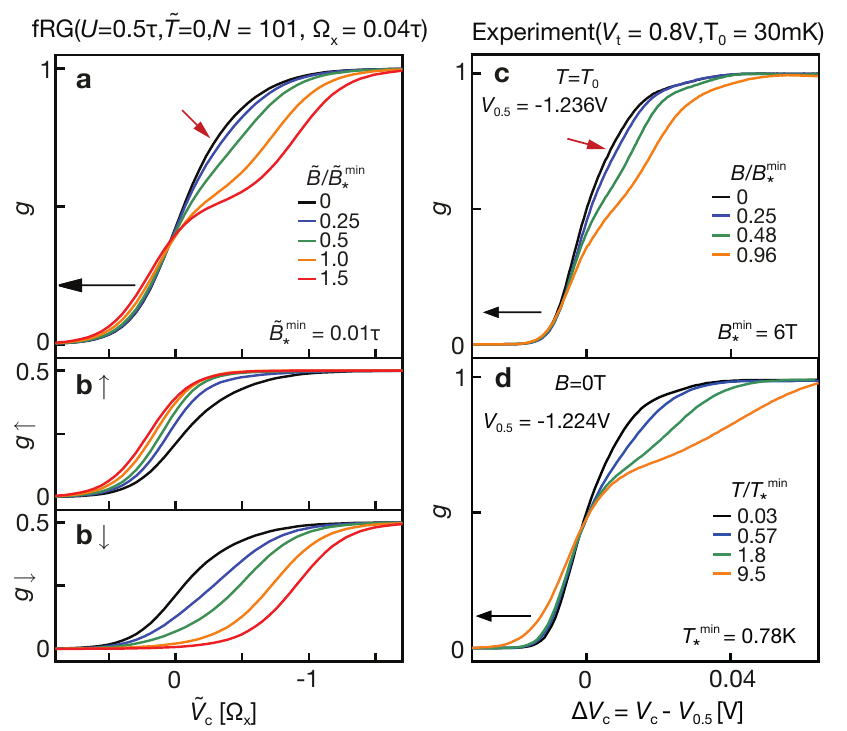}
\vspace{1mm}
\caption{\textbf{a}/\textbf{c}, Calculated/measured linear conductance $g(V_{\rm c})$
as a function of barrier height/gate voltage for several values of magnetic field, at
zero/low temperature we find good qualitative agreement: interactions cause a weak
shoulder even at zero field, which strengthens for intermediate fields and eventually
develops into a spin-resolved conductance step at high field.
\textbf{b$\uparrow$}$/$\textbf{b$\downarrow$}, Calculated spin-resolved conductance
curves for the same magnetic fields as in \textbf{a}. The conductance curves for
spin-up and spin-down react in an asymmetric fashion on an applied field: a
combination of  Pauli exclusion principle and Coulomb blockade (Hartree effect) leads
to a strong reduction of $\downarrow$-conductance, resulting in the phenomenon of the
0.7-anomaly. \textbf{d}, Measured conductance for several temperatures at zero field:
The 0.7-anomaly gets more pronounced with temperature, while all other features are
smeared out by thermal fluctuations.} \label{Fig:2}
\end{figure}

To investigate the influence of interactions we use the functional
Renormalization Group
(fRG)\cite{Wetterich1993,Metzner2011,Meden2002,Andergassen2004}. The fRG
approach in essence corresponds to an RG-enhanced perturbation theory in $U$
times the local density of states at the chemical potential. All results were
obtained for zero temperature, $T=0$.  The calculation was done in Matsubara-space, uses the
coupled-ladder approximation of the 2-particle vertex in both real-, and
frequency-space and is exact to second order in the interaction (see supplement
of \cite{Bauer_Nature_2013}).


Fig. 2a / Fig. 2c show the calculated/measured $\tilde{V}_{\rm c}$-dependence of the
linear conductance $g=G/G_Q$ of the lowest mode of a QPC for several values of magnetic field
and a finite interaction strength. We find very good qualitative agreement not only
for zero field, where the asymmetry of the step becomes manifest in a weak shoulder
 (marked by an arrow), but also at finite field, where the single step develops via a
 0.7-anomaly into a double step of width $g_{\rm{eff}}\mu_B B$. Fig. 2b$\uparrow$ and
 Fig. 2b$\downarrow$ show the calculated spin-resolved conductance for the same
 fields and interaction used in Fig. 2a. As expected, the conductance
 increases/decreases for the favoured/disfavoured (spin-up/spin-down) electrons. But
 unlike in the non-interacting case (not shown) the spin-down step is shifted much
 more strongly towards
 negative values of $\tilde{V}_{\rm c}$ than the spin-up shift is shifted towards
 positive values of $\tilde{V}_{\rm c}$. This can be explained as follows: Once a
 finite field breaks spin-symmetry, interactions push away spin-down electron out of
 the QPC's center, thereby depleting their density around the barriers top and
 consequently strongly reducing their probability of transmission. The 0.7-anomaly at
 finite magnetic field is a natural consequence of this interaction-induced
 asymmetry.

\begin{figure}[t]
\centering
\includegraphics[width=\linewidth]{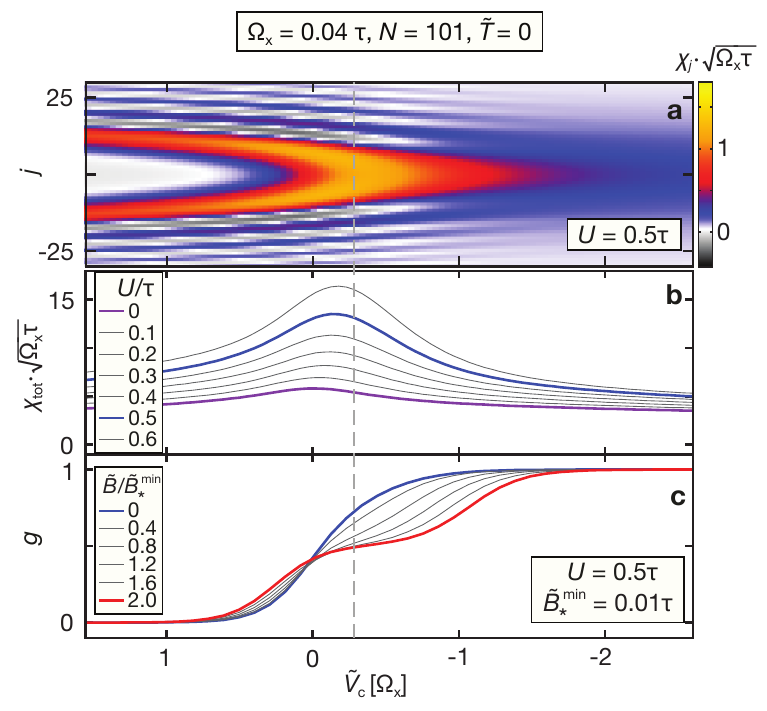}
\vspace{1mm}
\caption{\textbf{a}, Local spin susceptibility, $\chi_j(\tilde{V_{\rm c}})$, as a
function of site index $j$ and  barrier height $\tilde{V}_{\rm c}$ for a fixed value
of interaction strength, $U=0.5\tau$. \textbf{b}, The total spin-susceptibility of
the QPC, $\chi_{\rm tot}=\sum_j \chi_j$ for several values of interaction strength.
\textbf{a}. \textbf{c}, Calculated conductance curves as a function of barrier height
for several values of magnetic field. The strongest response of the system to a small
 applied magnetic field happens in the center of the barrier (see \textbf{a}) and
 coincides with the barrier height for which the 0.7-anomaly occurs (highlighted with
 the gray dashed vertical line around $\tilde{V}_{\rm c}=-0.25\Omega_x$).} \label{}
\end{figure}

As explained in detail in reference~\cite{Bauer_Nature_2013}, the origin of the
0.7-anomaly is caused by the presence in the LDOS of the van Hove ridge. Its apex
crosses the chemical potential, when the QPC is tuned into the sub-open regime, that
is, when the for conductance takes values $0.5\! \lesssim \! g \! \lesssim \!  1$. As
a consequence, the local spin-susceptibility, $\chi_j = \frac{1}{2}\left(\partial_h
m\right)_{h=0}$, shows not only a strong $j$-dependence due to the inhomogeneity of
the QPC, but also a strong $\tilde{V}_{\rm c}$-dependence, when the potential is
shifted through $\mu$ (see Fig. 3a). This also manifests itself in the total
spin-susceptibility of the QPC, $\chi_{\rm tot} = \sum_{j\in \rm{QPC}} \chi_j$, which
is plotted in Fig. 3b for several values of interaction strength. Three direct
consequences of interactions stand out: First, interactions strongly enhance the
effect of an applied magnetic field. Second, the maximum in the QPC's susceptibility
is shifted to somewhat lower values of $\tilde{V}_{\rm c}$ and, third, this maximum
occurs when the QPC is sub-open (gray dashed vertical line in Fig.3, compare with
conductance curves in Fig.  3c). These anomalous spatial structures in the spin
susceptibility serve as the main incentive for the experimental work described
further below, whose ultimate goal is to detect these structures by optical means.

Finally, we extracted the spin-splitting g-factor, $g_{\rm ss} = \frac{d \Delta E}{d
B}$, for several values of top gate voltage, $V_{\rm t}$.  Here $\Delta E$ depends on magnetic field and
is the energetic difference between spin-up and spin-down modes. It can be
extracted from the position of the maxima in the transconductance, $\frac{dg}{dV_c}$,
as a function of gate voltage, $V_c$, (see Fig. 4a/c), together with an appropriate
conversion factor $\Delta V_c = a \Delta E$. When varying the voltage of the top gate
we find a clear trend, namely that increasing $V_{\rm t}$ increases  $g_{\rm ss}$ as well (see
Fig. 4b). This can be explained as follows: The top gate in effect tunes the channel width; the more positive its voltage the narrower the channel
is, which in turn increases the effective interaction in the QPC. The prediction that
increased interaction strength causes a larger $g_{\rm ss}$ is confirmed by our
theoretical calculations, which reproduce the experimental trends quite beautifully
(see Fig. 4c and Fig. 4d).

\begin{figure}
\centering
\includegraphics[width=\linewidth]{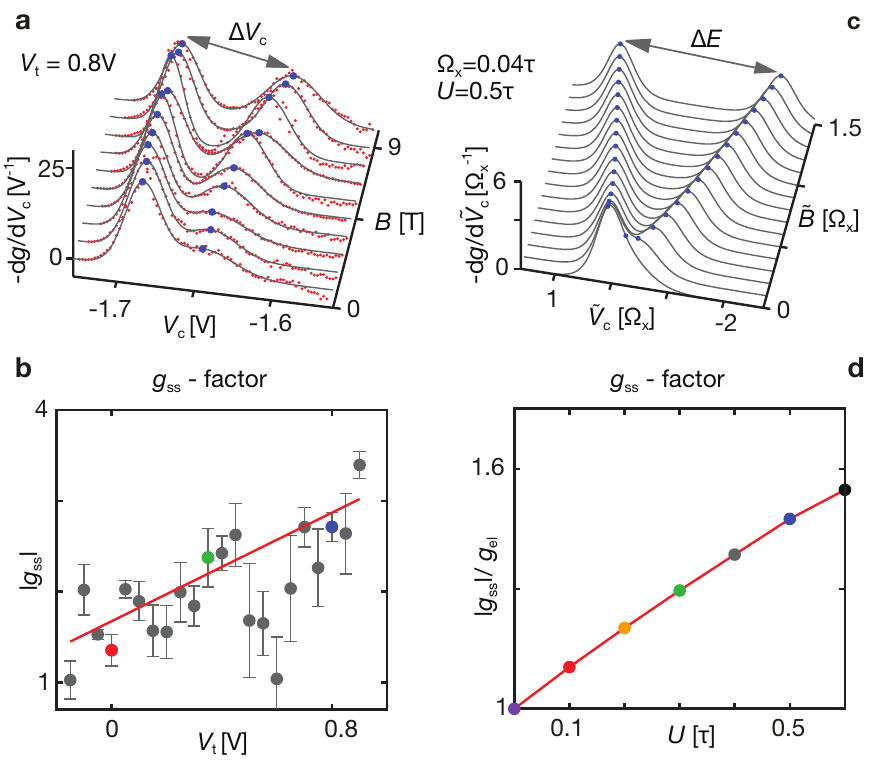}
\vspace{1mm}
\caption{\textbf{a}/\textbf{c}, Measured/calculated transconductance, $dg/dV_c$, at
fixed top gate/ interaction strength for several values of magnetic field.
\textbf{b},\textbf{d} Extracted values of the spin-splitting g-factor $g_{ss}$ for
several values of the top gate/interaction strength. The experiment confirms the
theoretical prediction that increasing interaction strength also increases the value
of $g_{ss}$.} \label{}
\end{figure}

\section{Theoretical motivation of the experiment}

Next, we describe ongoing experimental work, whose ultimate goal is to test the
following prediction emerging from the theoretical work described above:
For a QPC tuned in the regime of the 0.7-anomaly at zero external
magnetic field theory predicts an enhancement in the local
electron spin susceptibility \cite{Bauer_Nature_2013}. At finite magnetic
fields the enhanced spin susceptibility should give rise to
electron-spin polarization with a spatial distribution
characteristic of a QPC operated at the point of the 0.7-anomaly (see figure 3a).
Moreover, this polarization would also result in spin-sensitive
conductance. In principle, both signatures could be probed by
optical means: while spatially-resolved Kerr or Faraday rotation
could be used to map out the local spin-polarization in the
vicinity of the QPC, polarization-selective optical spin-injection
could be exploited to create an electron-spin imbalance across the
QPC to drive spin-polarized currents.

Our first step en-route to combined transport and optical
spectroscopy of a QPC in the 0.7-anomaly regime was to design a
heterostructure that would allow to implement both spin-sensitive
Faraday rotation and spin-selective charge carrier injection.

\vspace{\baselineskip} The following experimental part is divided into three sections. 
In Section~\ref{secSimu} we discuss the optimization process of 
the heterostructure design and the results of the simulations
performed with \textit{nextnano3}~\cite{nextNanoThree}. In
section~\ref{sec1stData} we present initial transport and optical
characterization measurements of the heterostructure. 
Section~\ref{sec8QPCsample} describes the present stage
of our experiments and provides perspectives for the combined
transport and optical spectroscopy of the 0.7-anomaly in a QPC.

\subsection{\label{secSimu}Semiconductor heterostructure design and simulations}

\begin{figure}[t]
\includegraphics*[width=\linewidth]{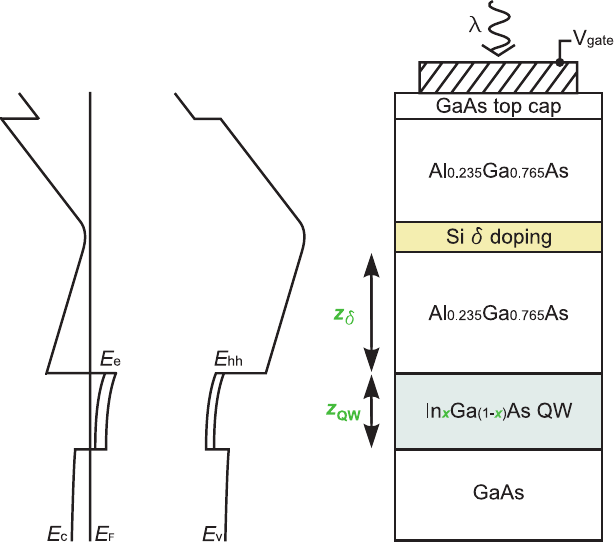}
\caption{Right: Schematic design of the heterostructure. A 2DES is
obtained in the InGaAs quantum well (blue) grown on top of GaAs by
electron transfer from the delta-doping region (yellow) within the
AlGaAs layer. A GaAs layer top cap layer terminates the
heterostructure. A semitransparent metal gate on top of the
heterostructure gives rise to a built-in Schottky potential and
allows to further bend the band structure via a voltage $V_{\rm gate}$.
The quantum well thickness $z_{QW}$, the indium concentration $x$
and the distance $z_{\delta}$ from the quantum well and the
delta-doping region were used as optimization parameters in 
simulations with \textit{nextnano3}. Left: Band structure profile
along the growth direction obtained from simulations for $V_{\rm
gate}=0$ . $E_{\rm c}$ and $E_{\rm v}$ denote the conduction and
valance band edges, $E_{\rm e}$ and $E_{\rm hh}$ the lowest
electron and heavy-hole levels confined in the quantum well, and
$E_{\rm F}$ is the Fermi energy, respectively.}
\label{heterostructure}
\end{figure}

The design of the heterostructure for combined 
transport and optical experiments was guided by two main objectives. On 
the one hand, we intended to realize a high quality two-dimensional 
electron system (2DES) suited for the observation of the 0.7-anomaly in a
QPC. On the other hand the sample structure should be designed to allow for
spin-selective optical excitations of charge carriers from the
valence band into the conduction band states of the 2DES just
above the Fermi level, and at the same time avoid excitations of
charge carriers in any other heterostructure layer. To make all
sample regions but the 2DES transparent to light at optical
frequencies that meet the resonance condition for interband
excitation of electrons into the Fermi sea we chose to embed an
In$_x$Ga$_{(1-x)}$As quantum well (QW) hosting the 2DES in higher
bandgap materials such as GaAs and AlGaAs. Accordingly, optical
excitations from the valence band states into the conduction band
states within the In$_x$Ga$_{(1-x)}$As QW exhibit the
smallest energy for interband transitions, provided that the
concentration $x$ of indium is finite. At the same time quantum
confinement associated with the QW removes the degeneracy of
heavy- and light-hole subbands at the $\Gamma$-point of bulk
zinc blende semiconductors, which in turn ensures "clean"
dipolar selection rules for spin-selective optical excitations
from the heavy hole subband at $E_{\rm hh}$ into the states at
$E_{\rm F}$ of the 2DES.

Fig.~\ref{heterostructure} illustrates the basic layout of our
heterostructure. The corresponding layer sequence along the sample
growth direction is shown in the right panel of
Fig.~\ref{heterostructure}. The In$_x$Ga$_{(1-x)}$As QW of
variable thickness $z_{QW}$ and an indium fraction $x$ in the
range of $0<x<0.1$ is sandwiched between GaAs and
Al$_{0.235}$Ga$_{0.765}$As that contains a delta-doping region
located at a distance $z_{\delta}$ above the QW. The AlGaAs layer
acts as a tunnelling barrier between the 2DES and the
semitransparent Schottky gate deposited on top of the
heterostructure. The overall thickness of the AlGaAs barrier was
set to half of the wavelength of the expected QW interband
transition to minimize optical interference effects. The silicon
delta-doping provides for excess electrons to form a 2DES inside
the QW and the final GaAs top cap layer prevents 
oxidization of the AlGaAs barrier. In the left panel of
Fig.~\ref{heterostructure} the corresponding band structure
profile calculated with \textit{nextnano3} is shown for zero external gate voltage, $V_{\rm gate}=0$,
and $x=0.07$, $z_{QW}=10{\rm nm}$ and $z_{\delta}=50{\rm nm}$. The band
profile bending is due to the built-in Schottky potential,
accounting for the lowest QW electron level $E_{\rm e}$ to lie below
the Fermi energy, in accord with our intention to create
a modulation-doped 2DES within the InGaAs QW. 

We recall the main properties of the intended heterostructure. 
First the QW containing the 2DES should exhibit the smallest 
interband transition energy with well defined dipolar selection
rules for spin-selective excitations. Second the semiconductor
matrix above and below the QW should be transparent at
the intended optical frequencies. Both criteria can be satisfied by 
the heterostructure layout of Fig.~1. Finally the density of the
2DES should be at least $2 \times 10^{11} {\rm cm}^{-2}$ to ensure the required transport
characteristics.

To this end we used \textit{nextnano3} to monitor the 2DES density
as a function of the optimization parameters $x$, $z_{QW}$ and
$z_{\delta}$. The objective was to achieve a maximum electron
density inside the QW of about $3 \times 10^{11}~{\rm cm}^{-2}$.
Simultaneously the interband transition wavelength of the QW
region, which follows from the energy difference between the lowest
QW hole level and the Fermi energy, was intended to lie above
$830~{\rm nm}$ in order to not overlap with optical transitions of
carbon impurities inherent to the molecular beam epitaxy (MBE) growth
process of the heterostructure. In Fig.~\ref{heterostructure}
these adjustable parameters for the simulations are highlighted in
green. Raising the QW thickness $z_{\rm QW}$ as well as the QW
indium content $x$ mainly increases the QW interband transition
wavelength. Reducing the distance $z_{\delta}$ between the QW and
the $\delta$-doping layer tends to increase the 2DES density.
However, at small QW thickness, proximity of the QW 2DES and the
doping layer, and a high indium concentration typically reduce
the mobility of the QW electrons and should be avoided.

\begin{figure}[t]
\includegraphics*[width=\linewidth]{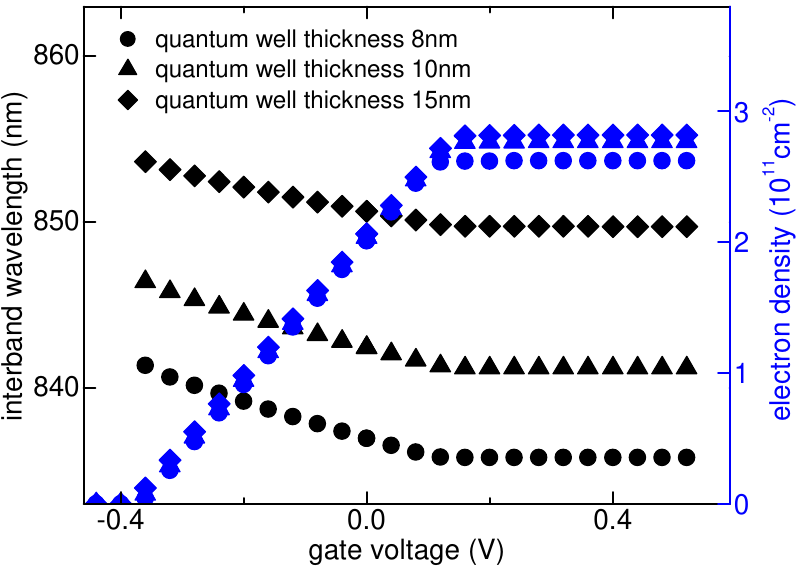}
\caption{Simulation results for the QW electron density (blue) and
the interband transition wavelength (black) as a function of gate
voltage. Results are shown for three different QW thicknesses of
$z_{\rm QW}=8~{\rm nm}$ (circles), $10~{\rm nm}$ (triangles) and
$15~{\rm nm}$ (squares) for a fixed indium content of $x = 0.07$ and a 
fixed spacer distance between the doping region and the QW of $z_{\delta}=50~{\rm
nm}$.} \label{n2DEGlambdaSIMU}
\end{figure}

Fig.~\ref{n2DEGlambdaSIMU} shows the simulation results for three
different heterostructures with an indium concentration of $x =
0.07$ and $z_{\rm QW}=50~{\rm nm}$. The QW thickness was taken as
$8~{\rm nm}$, $10~{\rm nm}$ and $~15~{\rm nm}$ to obtain a
variation in the QW electron density (blue) and the interband
transition wavelength (black) as a function of the voltage 
applied to the semitransparent top gate. Decreasing
the gate voltage increases the energy of the QW electron levels
with respect to the Fermi energy which gradually depletes the 2DES
density inside the QW. This depletion becomes increasingly
pronounced below gate voltages of $0.15~{\rm V}$ until the
pinch-off is reached at about $-0.4~{\rm V}$ for all three
heterostructures. The interband wavelength remains constant for
$V_{\rm gate}>0.25~{\rm V}$. At more negative gate voltages the
simulations predict a redshift of the resonance condition that is
associated with a decrease of the Fermi energy. In
Fig.~\ref{n2DEGlambdaSIMU} the maximum 2DES density as well as the
optical transition wavelength are close to 
our intended values.

\subsection{\label{sec1stData}Basic transport and optical characteristics}

Based on these simulation results a heterostructure was grown by
MBE with an indium concentration of $x = 0.07$, the separation
between the QW and the delta-doping layer of $z_{\rm QW}=50~{\rm
nm}$, and a QW thickness of $10~{\rm nm}$ (compare
Fig.~\ref{heterostructure}). Subsequently the sample material was
characterized with respect to basic transport and optical
properties. To determine the electron density and mobility of the
2DES a standard Hall bar geometry was used. The Hall bar
mesa was fabricated by conventional wet etching techniques and
AuGe/Ni/AuGe ohmic contacts were defined as Ohmic contacts to the
2DES. To allow control of the electron density a
semitransparent titanium gate with a thickness of $5~{\rm nm}$ was
deposited on top of the central region of the Hall bar structure.

\begin{figure}[t]
\includegraphics*[width=\linewidth]{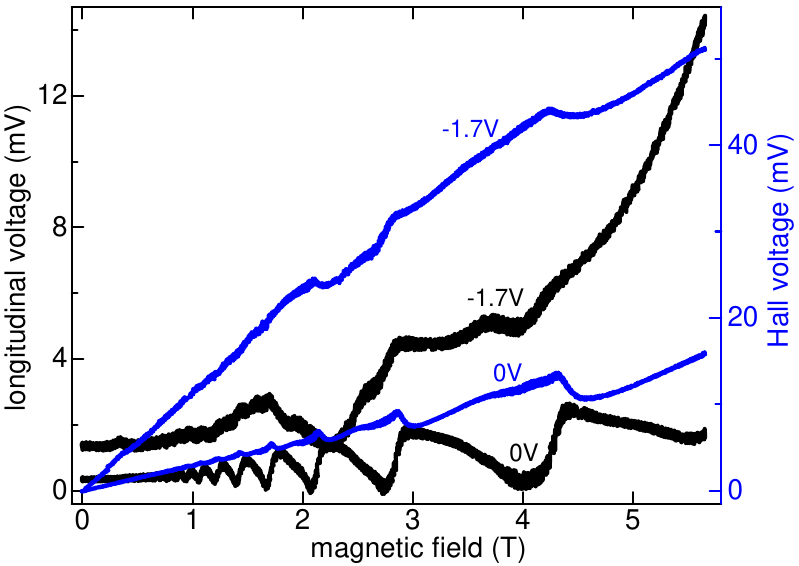}
\caption{Measured longitudinal (black) and Hall voltages (blue) as a function 
of the perpendicularly applied magnetic field at topgate voltages of $0{\rm V}$ 
and $-1.7{\rm V}$.} \label{LongHallVoltages}
\end{figure}

The electron density and mobility were extracted from
four-terminal dc Hall voltage measurements before the sample 
was subjected to light. Magnetic fields of up to
$5.7~{\rm T}$ were applied perpendicularly to the QW plane. By fitting
the Hall voltage $U_{\rm xy}$ versus the applied magnetic field
$B$ in the linear regime at low $B$ (Fig.~\ref{LongHallVoltages}) the carrier density of the 2DES 
is extracted by

\begin{equation}
\label{eqn2DESn}
n_{\rm 2DES} = \frac{I}{e \cdot dU_{\rm xy}/dB} .
\end{equation}
$e$ is  the elementary charge and $I$ is the current through the Hall bar. 
The mobility $\mu_{\rm 2DES}$ of the electron system
inside the QW was obtained from the longitudinal voltage
at zero magnetic field $U_{\rm xx}(B=0)$ (Fig.~\ref{LongHallVoltages}) using the relation

\begin{equation}
\label{eq2DESµ}
\mu_{\rm 2DES} = \frac{0.75}{e \cdot n_{\rm 2DES} \cdot U_{\rm xx}(B=0)}
\end{equation}
and the electron density $n_{\rm 2DES}$ obtained according to
Eq.~(\ref{eqn2DESn}). The number in the numerator is a scaling factor
imposed by the particular geometry of the employed Hall bar
structure. The same procedure was also carried out after
broad-band illumination of the sample.

In a second step we studied basic optical properties of the sample
by investigating the photoluminescence (PL) from the Hall bar. A
cryogenic confocal microscope with an optical spot size of $1~{\rm
\mu m}$ was used to record the local PL response, which was then
spectrally dispersed by a monochromator and detected with a
low-noise liquid nitrogen cooled CCD. All measurements were
carried out at a sample temperature of $4.2~{\rm K}$.

\begin{figure}[t]
\includegraphics*[width=\linewidth]{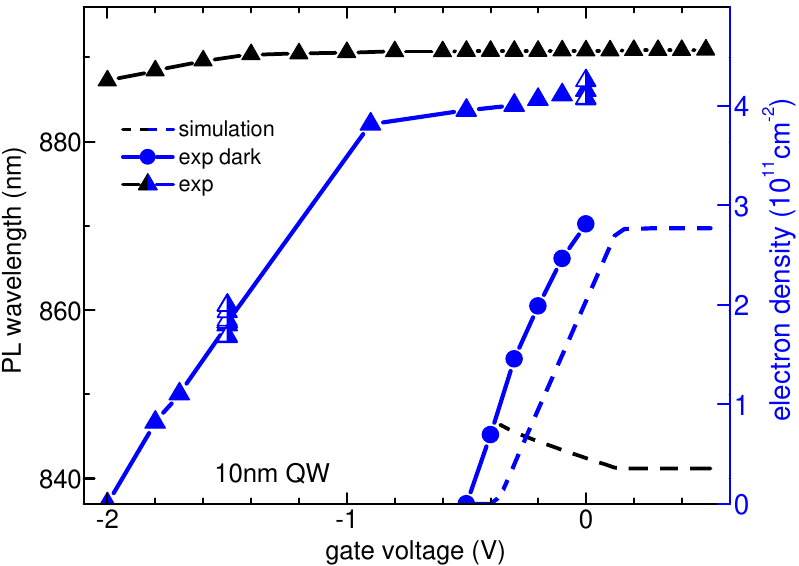}
\caption{Transport and optical characterization of the
heterostructure. Data (symbols; lines are guides to the eye) and
corresponding simulations as in Fig.~\ref{n2DEGlambdaSIMU} (dashed lines) 
for a heterostructure with a QW of thickness $z_{\rm QW}=10~{\rm nm}$, 7\% of indium concentration, and
$z_{\delta}=50~{\rm nm}$. The wavelength of the photoluminescence
peak maximum and the electron density of the QW 2DES are shown as
a function of gate voltage in black and blue, respectively. Circles and
triangles indicate the measurement results before and after light
illumination of the Hall bar, respectively. Half-filled triangles correspond to 
electron densities measured after repeated illumination of the sample. The photoluminescence
was obtained from the central region of the Hall bar using a
confocal setup with excitation powers of $7~{\rm \mu W}$ in the
range of $+0.6{\rm V}$ to $-1.0{\rm V}$ and  $0.3~{\rm \mu W}$
below $-1.0{\rm V}$ at an excitation wavelength of $830~{\rm nm}$.
The 2DES density was derived from standard Hall measurements. All
measurements were carried out at $4.2~{\rm K}$.} 
\label{n2DEGlambdaEXP}
\end{figure}

The combined transport and optical characterization results
are shown in Fig.~\ref{n2DEGlambdaEXP}. The 
interband transition wavelength (black) and the QW electron
density (blue) are shown as a function of gate voltage. Circles
(triangles) indicate the results of measurements done before
(after) illumination of the the sample with continuous wave (cw)
lasers (with $815~{\rm nm}$ and $830~{\rm nm}$ center wavelength).
Dashed lines show the corresponding simulation results from
Fig.~\ref{n2DEGlambdaSIMU} for comparison. In the simulations all
silicon dopants were assumed to be ionized, which is realized
experimentally by sample illumination. Despite an increase by $\sim 30\%$
of the 2DES density to around $4.2 \times 10^{11}{\rm cm}^{-2}$,
upon illumination, the simulated and experimental results are in very good agreement with the
predictions of the simulation. Consistently, the pinch-off gate 
voltage where the carrier density goes to zero is shifted to more negative values
upon illumination compared to the simulated pinch-off voltage. Repeated illumination
of the sample did not introduce further significant changes in the
2DES density (half-filled triangles in Fig.~\ref{n2DEGlambdaEXP}), indicating a long-term stability of the 2DES density
after the initial ionization of silicon dopants. The mobility of
the 2DES was determined to $\sim 70000~{\rm cm}^2/{\rm Vs}$ within
the entire gate voltage range above $-1.5{\rm V}$ after sample illumination 
(data not shown).

Despite good agreement between simulations and experiment for the
2DES density, we found considerable discrepancy between expected
and observed values for the wavelength of the optical transition
that we monitored via PL. Fig.~\ref{n2DEGlambdaEXP} shows the wavelength 
of the PL peak as a function of gate voltage recorded near the center of the Hall bar. Incident laser
excitation powers were $7~{\rm \mu W}$ in the voltage range
between $+0.6~{\rm V}$ and $-1.0~{\rm V}$ and $0.3~{\rm \mu W}$
below $-1.0~{\rm V}$, respectively. The excitation wavelength was
set to $830~{\rm nm}$, close to the wavelength region of carbon
impurity states in GaAs at $4.2~{\rm K}$. The mean difference between the simulated and the measured optical transition wavelength 
is about $50~{\rm nm}$ featuring different signs of the optical resonance shift. We speculate that the discrepancy
partially arises from excitonic effects and the quantum confined
Stark effect that were not accounted for in our simulations. Nevertheless, our main objective of the heterostructure design aiming at optical QW
transition energies below the band gap of GaAs was
successfully achieved.

\begin{figure}[t]
\includegraphics*[width=\linewidth]{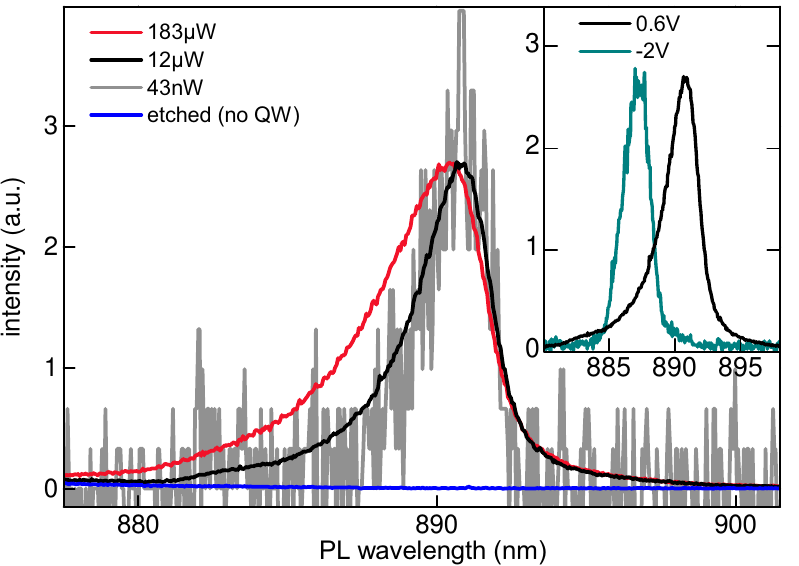}
\caption{Photoluminescence spectra recorded for a Hall bar sample
at $4.2{\rm K}$ with a QW thickness of $10~{\rm nm}$ and incident
excitation powers of $P_{\rm exc}=183~{\rm \mu W}$ (red), $12~{\rm
\mu W}$ (black, main and inset graph) and $43~{\rm nW}$ (grey)
scaled to maximum intensity values. The photoluminescence spectra
were measured in the central Hall bar region at gate voltages of
$V_{\rm gate}=+0.6~{\rm V}$ (flatband) under cw excitation at a
wavelength of $830~{\rm nm}$. The photoluminescence from an area
where the QW was etched away is shown in blue for reference.
Inset: photoluminescence spectra at two different gate voltages of
$+0.6~{\rm V}$ (black) and $-2.0~{\rm V}$ (green) for incident
excitation power of $300~{\rm nW}$.} \label{site+power}
\end{figure}

Fig.~\ref{site+power} shows the spectral characteristics of the
PL. The spectra were measured with a confocal setup in the central
Hall bar region at gate voltages of $+0.6~{\rm V}$ (flatband
condition) for a cw laser excitation wavelength of $830~{\rm nm}$
at $4.2K$. The PL exhibits an asymmetric profile reminiscent of
Fermi edge singularity \cite{Mahan_PR_1967,Hawrylak_PRB_1991,Kane_PRB_1994} 
even at lowest excitation powers down to $\sim 40~{\rm nW}$ (grey spectrum in
Fig.~\ref{site+power}). Unambiguously, the source of the PL is the
QW, since no PL was detected in the relevant spectral window from
sample regions were the QW was etched away (blue spectrum in
Fig.~\ref{site+power}). We find indications of higher-energy
shoulders at $883~{\rm nm}$ and $885~{\rm nm}$ that emerge with
increasing excitation powers accompanied by a blue-shift of the PL
maximum. These characteristics were consistently found at
different spatial locations of the Hall bar structure were the QW
was not etched away. We also found that the PL was sensitive to
the gate voltage. The inset of Fig.~\ref{site+power} compares the
PL spectra at $V_{\rm gate}=+0.6~{\rm V}$ and $-2.0~{\rm V}$,
showing a clear blue-shift of the PL resonance with more negative
gates voltages that was accompanied by a gradual evolution of the
PL line shape towards a symmetric Gaussian peak (fit not shown).

\subsection{\label{sec8QPCsample}Outlook}

\begin{figure}[t]
\includegraphics*[width=\linewidth]{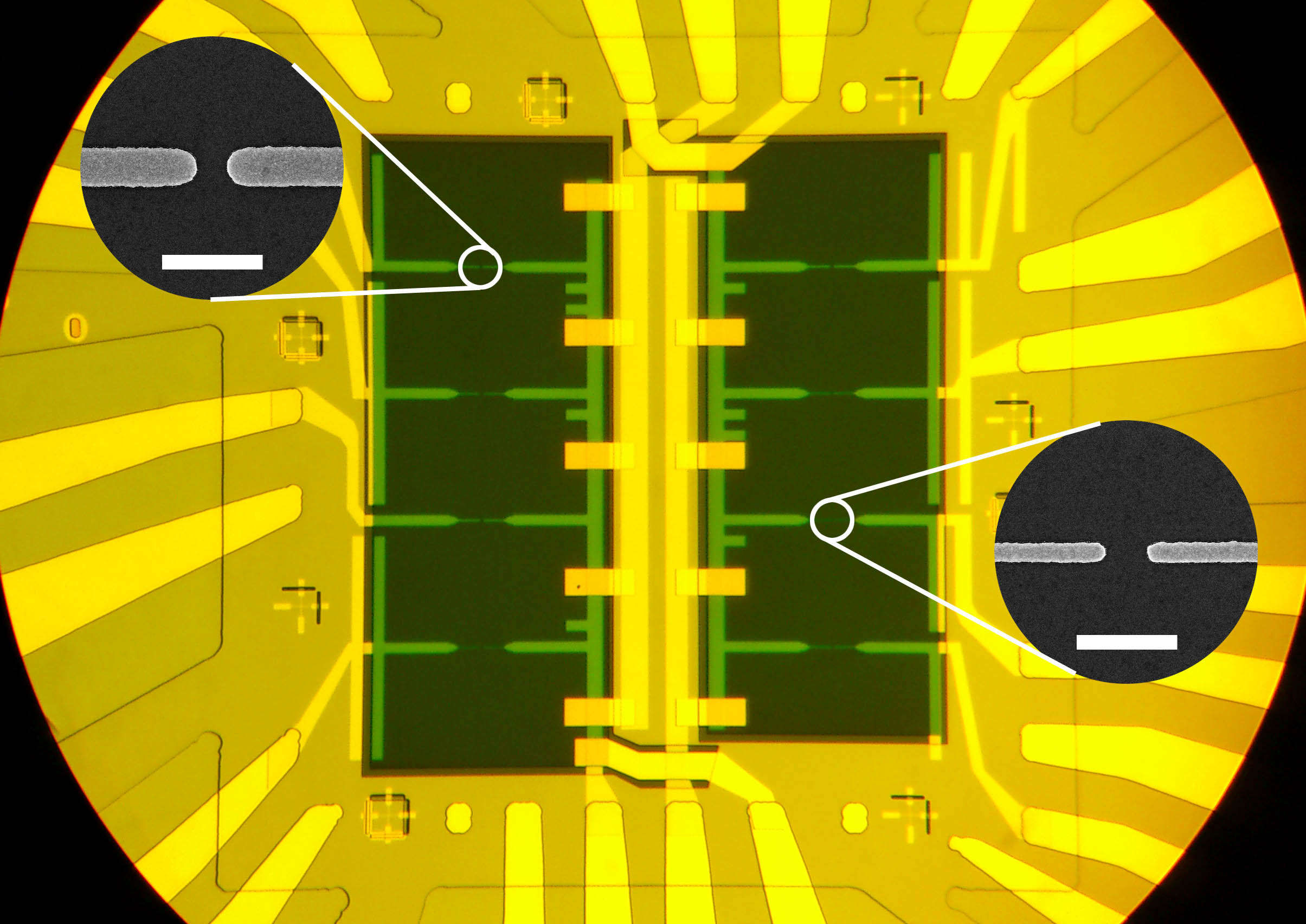}
\caption{Optical microscope image of the sample layout with eight
QPCs and a global top gate fabricated on the $10~{\rm nm}$ InGaAs
QW heterostructure. Optical and electron-beam lithography followed
by gold deposition and lift-off were used to define gates on top
of a square mesa with an edge length of $160~{\rm \mu m}$. The
QPCs are formed between the ends of finger-like gates (also shown
as scanning-electron micrograph insets each including a horizontal scale 
bar corresponding to $1{\rm \mu m}$ length) of different geometries. On the top of the mesa a
semitransparent gate was deposited that is electrically
disconnected by cross-linked PMMA from all other gates used to
define the QPCs.} \label{8QPCsample}
\end{figure}

The basic properties of the heterostructure described above
represent a promising starting point for in-detail transport and
optical studies of the 0.7-anomaly in QPCs. Fig.~\ref{8QPCsample}
shows an optical micrograph of our present sample layout
implemented on a heterostructure that contains a 2DES hosted by an
InGaAs QW of $10~{\rm nm}$ thickness. Gold gates
defined by optical lithography (outer yellow pads) connect to
inner Au gates processed by electron beam lithography (light
yellow) across the mesa-edges (centered square and starlike
surrounding connections). Eight QPCs of different widths and lengths
of the gated constrictions between $200{\rm nm}$ and $500{\rm nm}$ 
are covered by layers of cross-linked Poly 
Methyl methacrylate (PMMA) (dark grey). The latter electrically isolates the QPC 
gates from the two semitransparent
Nickel-Chromium top gates of $5{\rm nm}$ thickness (black
rectangles on top of the PMMA in Fig.~\ref{8QPCsample}). They 
allow to simultaneously having optical access to the 2DES
layer and being able to tune its carrier density. The insets in
Fig.~\ref{8QPCsample} show SEM pictures of two specific QPC
geometries.

The next step will involve transport experiments to study the $0.7$-anomaly 
as a function of the QPC geometries in the accessible experimental parameter space. 
In opto-transport experiments we will then attempt to optically monitor the field-dependence of the spin-up 
and spin-down densities in the vicinity of the QPC as a function of QPC gate voltage $V_{\rm c}$ and top gate voltage 
$V_{\rm t}$. We will also aim to perform near-resonant injection of spin-polarized electrons in the 
vicinity of a QPC to observe spin-selective transport. We
intend to exploit the full potential of the combined optical and transport setup in terms
of position, energy and spin selective spectroscopy to shed light on
the microscopic origin of the 0.7-anomaly.

\begin{acknowledgement}
We acknowledge discussions with J.~P.~Kotthaus. We
gratefully acknowledge funding by the Deutsche
Forschungsgemeinschaft within the priority program "Semiconductor
Spintronics" (SPP 1285) and the center of Excellence, Nanosystems Initiative Munich (NIM), and financial support from the 
Center for NanoScience (CeNS).
\end{acknowledgement}




\end{document}